\documentclass[aps,prl,twocolumn,superscriptaddress,floatfix,sort&compress,longnamesfirst]{revtex4-2}

\usepackage{amsmath,amssymb}
\usepackage{bm,bbm}
\usepackage{hyperref}
\usepackage{dcolumn}
\hypersetup{
	colorlinks=true,
	linkcolor=blue,
	filecolor=blue,
	urlcolor=blue,
	citecolor=blue,
}
\usepackage{graphicx}

\begin{document}

\title{Open-system analogy of Berry conjecture}

\author{Yaohua Li}
\email{liyaohua20@mails.tsinghua.edu.cn}
\affiliation{State Key Laboratory of Low-Dimensional Quantum Physics, Department of Physics, Tsinghua University, Beijing 100084, P. R. China}
\author{Yunhan Wang}
\affiliation{State Key Laboratory of Low-Dimensional Quantum Physics, Department of Physics, Tsinghua University, Beijing 100084, P. R. China}
\author{Yong-Chun Liu}
\email{ycliu@tsinghua.edu.cn}
\affiliation{State Key Laboratory of Low-Dimensional Quantum Physics, Department of Physics, Tsinghua University, Beijing 100084, P. R. China}
\affiliation{Frontier Science Center for Quantum Information, Beijing 100084, China}

\date{\today}

\begin{abstract}
	Berry conjecture is central to understanding quantum chaos in isolated systems and foundational for the eigenstate thermalization hypothesis. Here we establish an open-system analogy of the Berry conjecture, connecting quantum steady states to classical dissipative attractors in the semiclassical limit. We demonstrate that the Wigner distribution of quantum steady states delocalizes over classical chaotic attractors in the semiclassical limit. We validate this correspondence using a Floquet Kerr oscillator. In the chaotic phase, the quasi-steady state is dominated by the chaotic delocalization instead of the quantum fluctuations, resulting in entropy divergence in the semiclassical limit. This entropy divergence provides a robust chaos signature beyond non-Hermitian random matrix approaches. We further identify dissipative phase transitions via Liouvillian gap closures, revealing a discrete time crystal phase and its breakdown into chaos at strong driving. Our framework thus establishes a universal paradigm for quantum chaos in open systems.
\end{abstract}
\maketitle

{\it Introduction.}---Chaos is a phenomenon of exponential sensitivity of infinitely small perturbations in the initial conditions \cite{ueda_randomly_1979,thompson_nonlinear_2002}. The exponential sensitivity, as characterized by the Lyapunov exponents, can result in unpredictable and irreversible dynamics in deterministic but nonintegrable systems. Moreover, a chaotic system is ergodic in the possible phase space, which provides an important understanding of the ergodicity in thermodynamics \cite{deutsch_eigenstate_2018}.

The generalization of classical chaos to quantum systems is not straightforward, because it is difficult to define both the quantum trajectory and the distance between two states \cite{dalessio_quantum_2016}. Fortunately, a numerical correspondence between random matrices and isolated quantum systems with chaotic classical counterparts has been discovered. It leads to the quantum distinction of quantum chaos based on the statistical properties of the quantum spectra. This method is concluded into two conjectures: the Berry-Tabor conjecture \cite{berry_level_1977} and the Bohigas-Giannoni-Schmit conjecture \cite{bohigas_characterization_1984}. Based on the two conjectures, the level statistics of regular and chaotic systems can be predicted by randomly distributed energy levels and the eigenvalues of Gaussian random matrices, respectively \cite{atas_distribution_2013,tian_statistics_2024}. There are also great progresses in analytical derivation aiming to relate the random matrix spectra with the quantum chaotic systems, including for single-particle \cite{berry_semiclassical_1997,muller_semiclassical_2004} and many-body systems \cite{kos_many-body_2018}.

For classical systems with dissipation, the nonintegrable chaotic system exhibits strange attractor, distinct from regular attractors such as fixed points and periodic orbits in integrable systems \cite{szlachetka_chaos_1993}. The chaotic dissipative dynamics have been discovered in various quantum systems, including the Dicke model \cite{stitely_nonlinear_2020,villasenor_analysis_2024,mondal_transient_2025}, spin chains \cite{akemann_universal_2019,xu_accessing_2020}, and optomechanical systems \cite{monifi_optomechanically_2016,navarro-urrios_nonlinear_2017,zhu_cavity_2023}. There is also a non-Hermitian random matrix theory for the distinction of integrable and nonintegrable dissipative systems, as known as the Grobe-Haake-Sommers (GHS) conjecture \cite{grobe_quantum_1988,grobe_universality_1989}. It relates the spectra of the Liouvillian superoperator in the semiclassical limit to the spectra of the non-Hermitian random matrix \cite{sa_complex_2020,li_spectral_2021,kawabata_symmetry_2023}. But the Liouvillian spectra is far more difficult to be obtained than the Hamiltonian spectra when approaching the semiclassical limit, due to the quadratic dimensions. Moreover, a counterexample based on the open Dicke model showed that the GHS conjecture may not be universal in open quantum systems \cite{villasenor_breakdown_2024}.

Simultaneously with the Berry-Tabor conjecture, Berry also discovered a remarkable connection between the quantum quasi-probability distribution in the semiclassical limit and the classical distribution of chaotic isolated systems, as known as the Berry conjecture \cite{berry_regular_1977}. This conjecture states that for an $N$-partical chaotic system, in the limit $\hbar\to0$, the local average of Wigner distribution $W(\mathbf{x}, \mathbf{p})$ of energy eigenstates reduces to the microcanonical distribution:
\begin{equation}
	\overline{W(\mathbf{x}, \mathbf{p})} = \frac{\delta\left[E-H(\mathbf{x}, \mathbf{p})\right]}{\int d\mathbf{x} d\mathbf{p}\delta\left[E-H(\mathbf{x}, \mathbf{p})\right]},
\end{equation}
where $\overline{W}$ represents the local average of the Wigner distribution, $\mathbf{x}$, $\mathbf{p}$ are the $N$-dimensional coordinates and momenta, $E$, $H$ are the energy and classical Hamiltonian, and $\delta[\cdots]$ is a one-dimensional Dirac delta function. This property has been proven to be a sufficient condition for the emergence of the eigenstate thermalization hypothesis (ETH), which bridges the microscopic reversibility of quantum mechanics and the macroscopic irreversibility of thermodynamics \cite{srednicki_chaos_1994}. Despite the great success in isolated systems, the direct questions are whether we can establish a semiclassical connection between the quasi-probability distribution of open quantum systems and classical distribution when the corresponding classical systems are chaotic, and what the connection can predict.

Here we show such a connection can be established between the quasi-steady states and dissipative attractors in the Poincar{\' e} section. We develop a general framework in analogy to the Berry conjecture in isolated systems. While a regular distribution is delocalized due to quantum fluctuations (or thermal fluctuations), the Wigner distribution in the chaotic region is further delocalized due to the ergodicity of chaos. In the semiclassical limit, the quantum fluctuations can be neglected compared to the chaotic delocalization, and the quantum chaotic distribution approaches the classical chaotic distribution.

We corroborate our theory through an example of Floquet Kerr oscillator. When increasing the nonlinear Floquet driving strength, the system will first enter a dissipative discrete time crystal phase before the quantum chaos emerges. The fluctuations remain limited in the regular phases and diverge in the chaotic phase. We show the quantum-classical correspondence by approaching the semiclassical limit. Moreover, by computing the quantum entropy of the quasi-steady states, we obtain clear numerical evidence of chaotic delocalization and show that it can be used to distinguish regular and chaotic open quantum systems. As the steady state is much easier to be obtained than the full Liouvillian spectra, this method can be applied to more complicated systems beyond the non-Hermitian random matrix theory. Finally, we further determine the quantum phase boundaries based the closing of Liouvillian gap, which coincide with the mean-field prediction and the results of steady observable.

{\it General theoretical framework.}---
For integrable isolated systems, the Wigner distribution of a quantum state is around the corresponding classical orbit with quantum fluctuations. In the semiclassical limit $\hbar\to0$, the quantum fluctuations disappear, and the average Wigner distribution reduces to the classical orbit. Based on the Liouville theorem and the ergodicity of the chaotic system, a similar connection for nonintegrable systems leads to the Berry conjecture, i.e., the average Wigner distribution reduces to the microcanonical distribution in the semiclassical limit.

\begin{figure}[t]
	\centering
	\includegraphics[width=0.49\textwidth]{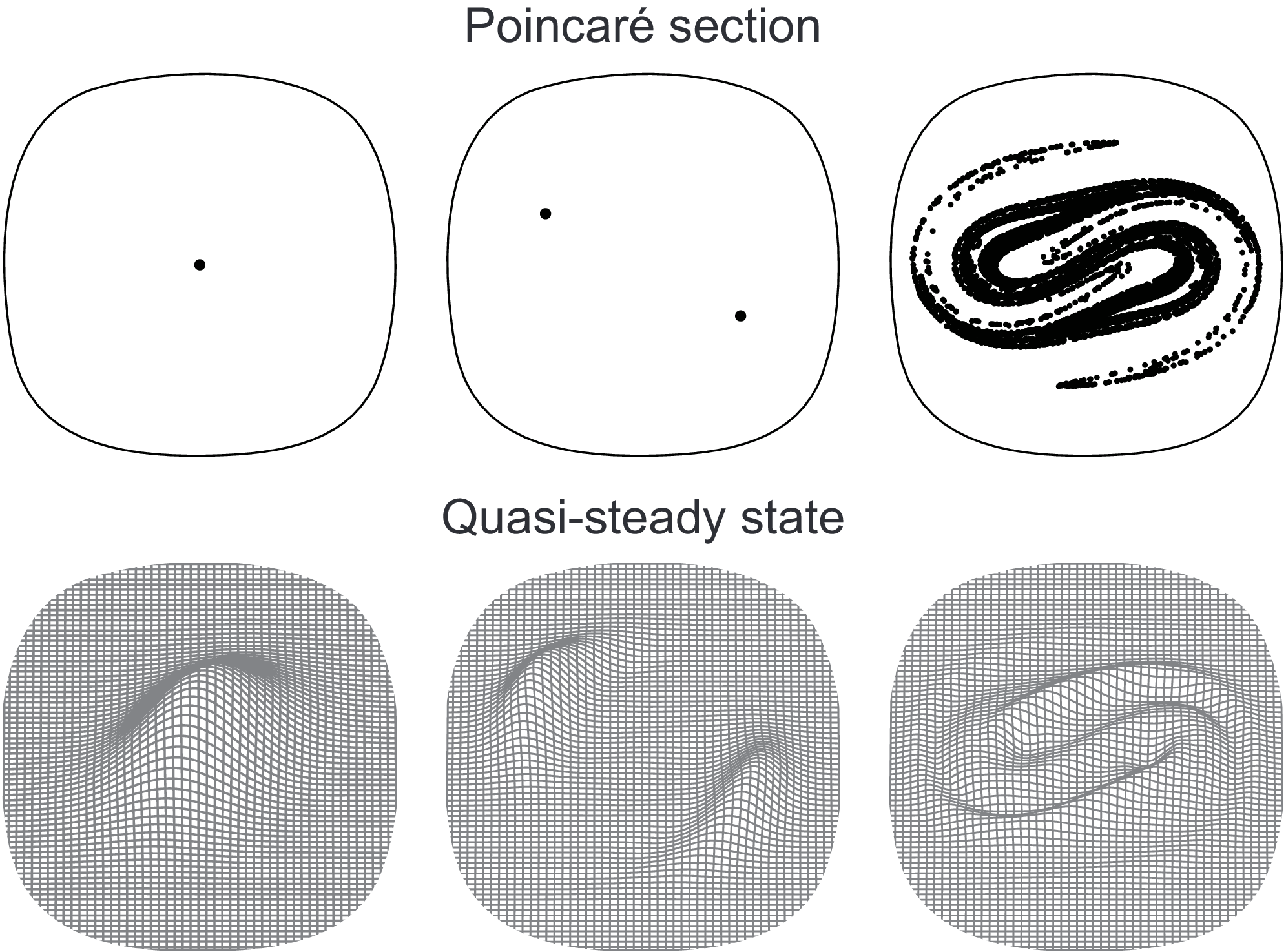}
	\caption{Schematic of quantum-classical correspondence of the chaotic distribution. The first row contains three attractors in the Poincar{\' e} section: single fixed point, double fixed points, and dissipative chaos. The second row is Wigner distributions of the corresponding quasi-steady states.}
	\label{fig:a}
\end{figure}

For isolated systems, the semiclassical theories such as Wenzel-Kramers-Brillouin (WKB) approximation expand the results as a power series in $\hbar$. Consequently, the semiclassical limit naturally arises as $\hbar\to0$. For open systems, it is difficult to obtain similar expansion. However, there is another way to investigate the quantum-classical crossover by tuning the excitation number $N$. The semiclassical limit is achieved by letting $N\to\infty$.

As illustrated in Fig. \ref{fig:a}, the quantum distribution of quasi-steady states in an open system is approximately distributed around the corresponding mean-field attractors \cite{dutta_quantum_2025}. This is exact in the semiclassical limit and depends on the accuracy of the mean-field approximation. Importantly, this behavior is not only satisfied by the regular attractor but also for the chaotic attractor, as verified by quantum state diffusion theory \cite{sup,plenio_quantum-jump_1998,spiller_emergence_1994,iwaniszewski_chaos_1995,brun_quantum_1996,adamyan_chaos_2001,dahan_classical_2022,mondal_dissipative_2025}.
For regular attractors, the quantum fluctuations from system-environment coupling are the only origin of the delocalization in the phase space, which can be neglected compared to the large excitation in the semiclassical limit. However, in the chaotic region, the major origin of the delocalization is the delocalized chaotic attractor, which dominates in the quantum-classical crossover when the quantum fluctuations can be neglected. As an analogy to the Berry conjecture, this connection can be summarized as an equation that holds in the semiclassical limit:
\begin{equation}
	\overline{W(\alpha, \alpha^{\ast})} = \frac{\delta(\alpha, \Sigma)}{\int d^{2}\alpha \delta(\alpha, \Sigma)},
\end{equation}
where $\Sigma$ represents the area of the chaotic attractor. $\delta(\alpha, \Sigma)$ is a two-dimensional delta-like function that is equal to 1 for $\alpha\in\Sigma$ and equal to 0 for $\alpha\notin\Sigma$.
\begin{table*}
	\caption{\label{tab:table1}%
	Nonexhaustive comparison between Berry conjecture and its analogy in open quantum systems.}
	\begin{ruledtabular}
	\begin{tabular}{ccc}
		& Berry conjecture & open-system analogy \\ \hline
		classical phase space & energy surface & dissipative attractor \\
		quantum state & eigenstate & steady state \\
		Wigner distribution & microcanonical distribution & delocalized distribution \\
		semiclassical description & e.g. WKB approximation & mean-field approximation \\
		semiclassical limit & $\hbar\to0$ & $N\to\infty$ \\
		thermalization & ETH & limited \\
	\end{tabular}
	\end{ruledtabular}
\end{table*}

Besides the analogies described so far [cf. Table \ref{tab:table1}], the major difference is the relation to quantum thermalization. The nonintegrability is necessary for quantum thermalization in isolated systems but is not in open systems \cite{ashida_thermalization_2018}. Moreover, the dissipative attractors in an open system have zero measure in phase space, unlike the energy surface in an isolated system with invariant phase-space density. The ergodicity in the energy space leads to a microcanonical distribution and means the emergence of quantum thermalization, but the ergodicity of a chaotic attractor can not provide a similar inference, except for some special cases \cite{shirai_thermalization_2020,ferrari_chaos_2025}.

{\it Model and quantum-classical connection.}---
In the following, we consider a nonlinear bosonic mode with Kerr nonlinearity ($U$), and Floquet squeezing drive ($\varepsilon$). The Hamiltonian reads
\begin{equation}
	H(t) = -\Delta a^{\dag}a + \frac{U}{2}a^{\dag2}a^{2} + \frac{\varepsilon(t)}{2} a^{\dag2} + \frac{\varepsilon(t)^{\ast}}{2}a^{2},
\end{equation}
where $a$ is the annihilation operator and $\Delta$ is the detuning between the driving field and the bosonic mode. We assume a two-step drive $\varepsilon(t)=\varepsilon_{1}$ or $\varepsilon_{2}$ with equal time $T/2$,
where $T$ is the period. Here we mainly consider a special case for $\varepsilon_{2}=-\varepsilon_{1}=\varepsilon_{0}$. The general results can be found in the supplementary materials \cite{sup}. Then the density matrix of the system obeys a master equation of the form
\begin{equation}
	\dot{\rho}(t)=i[\rho(t), H(t)]+\kappa \mathcal{D}(a)\rho(t)\equiv\mathcal{L}(t)\rho(t),
\end{equation}
where $\kappa$ is the dissipation rate, $\mathcal{D}(a)\rho=a\rho a^{\dag}-(a^{\dag}a\rho+\rho a^{\dag}a)/2$ is the Lindblad dissipator, and $\mathcal{L}(t)$ denotes the time-dependent Liouville superoperator.

\begin{figure}[b]
	\centering
	\includegraphics[width=0.49\textwidth]{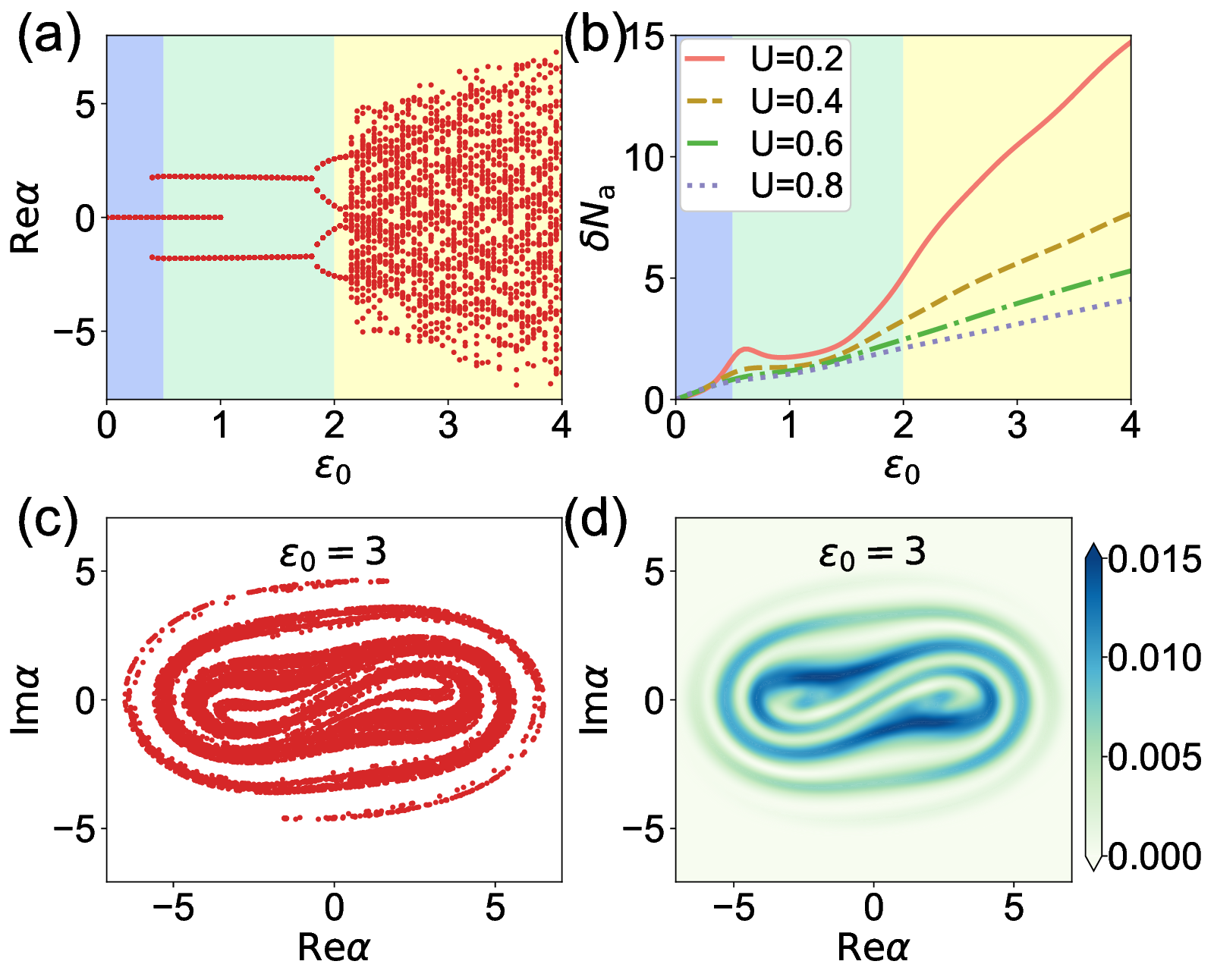}
	\caption{Classical chaotic attractor and chaotic quasi-steady state. (a),(b) Real part of classical attractor (a) and boson number fluctuations ({b}) as a function of the Floquet driving strength $\varepsilon_{0}$. The three colored regions in \textbf{a} denote three quantum phases, including two regular phases and a chaotic phase. The approximate boundaries $\varepsilon\approx0.5, 2$ are determined by the Liouvillian gap. {(c)} Detail classical attractor for $\varepsilon_{0}=3$. {(d)} Wigner distributions of the quasi-steady states for $\varepsilon_{0}=3$. Other parameters are $\Delta=-1$, $U=0.2$, $\kappa=0.5$, and $T=2$.}
	\label{fig:b}
\end{figure}

Before starting the quantum derivation, we first analyze the classical dynamics and attractors of the mean-field amplitude $\alpha=\langle a \rangle$, which satisfies
\begin{equation}
	\dot{\alpha}(t)=\left(-\frac{\kappa}{2}+i\Delta-iU\left|\alpha(t)\right|^2\right)\alpha(t)-i\varepsilon(t) \alpha^{\ast}(t).
\end{equation}
For a periodically driven system, the dynamic structure is characterized by the attractors in the Poincar{\' e} section. As shown in Fig. \ref{fig:b}a, when increasing the Floquet driving strength, there are mainly three different phases with attractors of single fixed point, double fixed points, and chaos. As a comparison, we also show the three phases in color obtained in the quantum results (discussed later).

In the quantum region, the quasi-steady state is given by $\mathcal{L}_{\mathrm{eff}}\rho_{\mathrm{ss}}=0$, where the effective Liouvillian $\mathcal{L}_{\mathrm{eff}}$ satisfies \cite{schnell_is_2020,chen_periodically_2024,chen_engineering_2025}
\begin{gather}
	\mathcal{L}_{\mathrm{eff}}=\frac{1}{T}\ln \mathcal{U}(T),\\
	\mathcal{U}(T)=e^{\mathcal{L}_{2}T/2}e^{\mathcal{L}_{1}T/2}.
\end{gather}
In Fig. \ref{fig:b}b, we show one of the main results of the boson number fluctuation that supports our theory. The quantum fluctuations remain fixed in the regular phases and diverge in the chaotic phase. The divergent fluctuation originates from the chaotic delocalization and allows us to neglect the fixed quantum fluctuations from the environment. Figure \ref{fig:b}c and \ref{fig:b}d are representative chaotic attractors and the corresponding quantum distribution of the quasi-steady state, respectively. They exhibit perfect correspondence, even not very close to the semiclassical limit \cite{sup}.

{\it Delocalization and entropy.}---
To quantitatively analyze the chaotic delocalization, it is useful to compare the entropy between the quasi-steady state and a bosonic thermal state. The van Neumann entropy $S=-\rho\ln\rho$ of a thermal state  is given by
\begin{gather}
	S_{0, \mathrm{v}}=(N_{\mathrm{a}}+1)\ln(N_{\mathrm{a}}+1)-N_{\mathrm{a}}\ln N_{\mathrm{a}},
\end{gather}
which is a function of the average boson number $N_{\mathrm{a}}$.

\begin{figure}[t]
	\centering
	\includegraphics[width=0.49\textwidth]{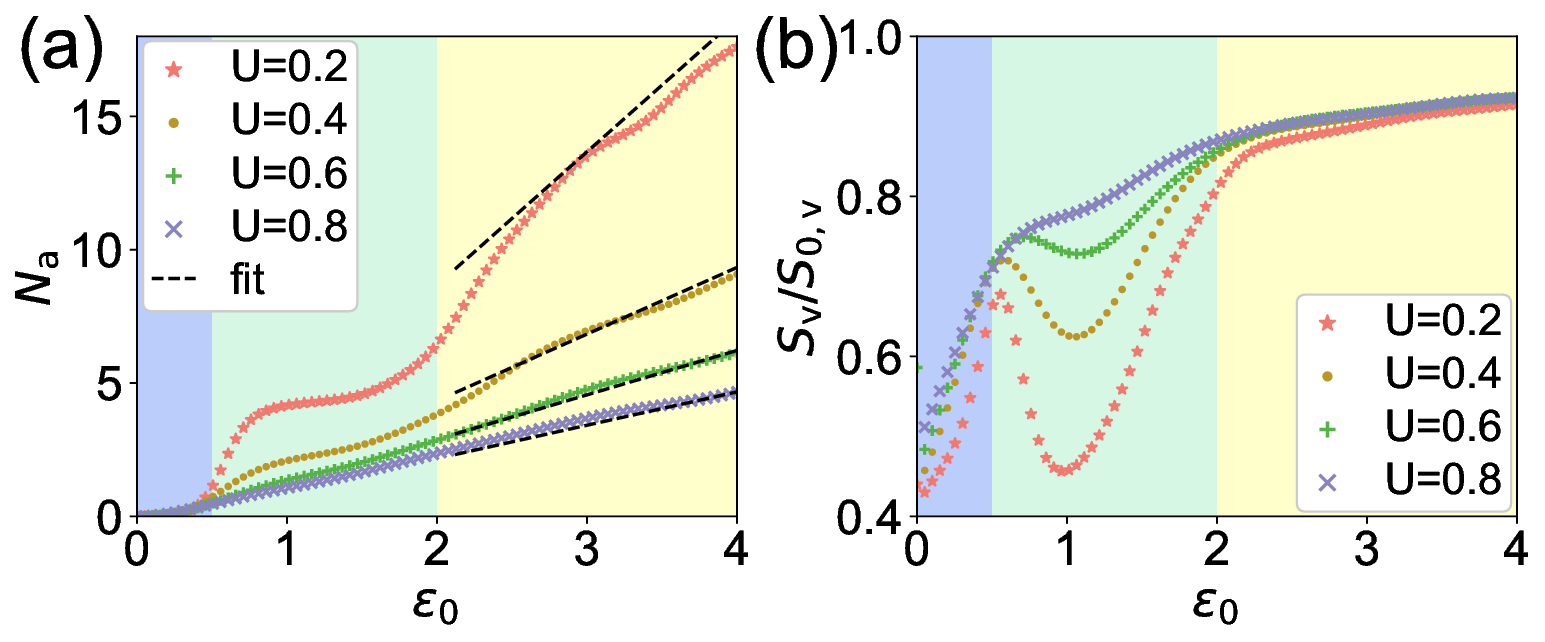}
	\caption{{Quasi-steady observables and comparison with the thermal equilibrium state.} {(a),(b)} The average boson number $N_{\mathrm{a}}$ ({a}) and the von Neumann entropy $S_{\mathrm{v}}$ ({b}) of the quasi-steady state as a function of the Floquet driving strength $\varepsilon_{0}$. The black dashed lines in ({a}) are obtained by numerical fitting through a function of $N_{\mathrm{a}}\approx (\varepsilon_{0}-\varepsilon_{\mathrm{c}})/U$, and we obtain $\varepsilon_{\mathrm{c}}\approx0.27$. The colored lines in {(a) and (b)} represent different Kerr nonlinearity strength $U$. Other parameters are the same as Fig. \ref{fig:b}.}
	\label{fig:c}
\end{figure}

In Figs. \ref{fig:c}{a} and \ref{fig:c}b, we plot the average boson number $N_{\mathrm{a}}$ and the ratio of the van Neumann entropy $S_{\mathrm{v}}/S_{0, \mathrm{v}}$ between the quasi-steady state and the thermal state, respectively. The average boson number is near zero in the single-fixed-point phase and nearly invariant in the double-fixed-point phase, similar to the behaviors of mean-field regular attractors [see Fig. \ref{fig:b}a]. Differently, in the chaotic phase, the average bosonic number approximately satisfies $N_{\mathrm{a}}\approx (\varepsilon_{0}-\varepsilon_{\mathrm{c}})/U$. The black dashed lines in Fig. \ref{fig:c}{a} are numerical fitting results, which give $\varepsilon_{\mathrm{c}}\approx0.27$. As shown in Fig. \ref{fig:c}b, the ratios of the von Neumann entropy are away from 1 in the fixed-point phases and remain near 1 in the chaotic phase when approaching the semiclassical limit. In the fixed-point phases, external drive breaks the thermal equilibrium and results in non-equilibrium quasi-steady states. The more we approach the semiclassical limit $U\to0$, the farther the quasi-steady state is away from equilibrium. It strongly suppresses the entropy in the quasi-steady state. Differently, the delocalized chaotic quasi-steady state possesses divergent entropy in the semiclassical limit with the same scaling as the thermal state.

\begin{figure}[t]
	\centering
	\includegraphics[width=0.49\textwidth]{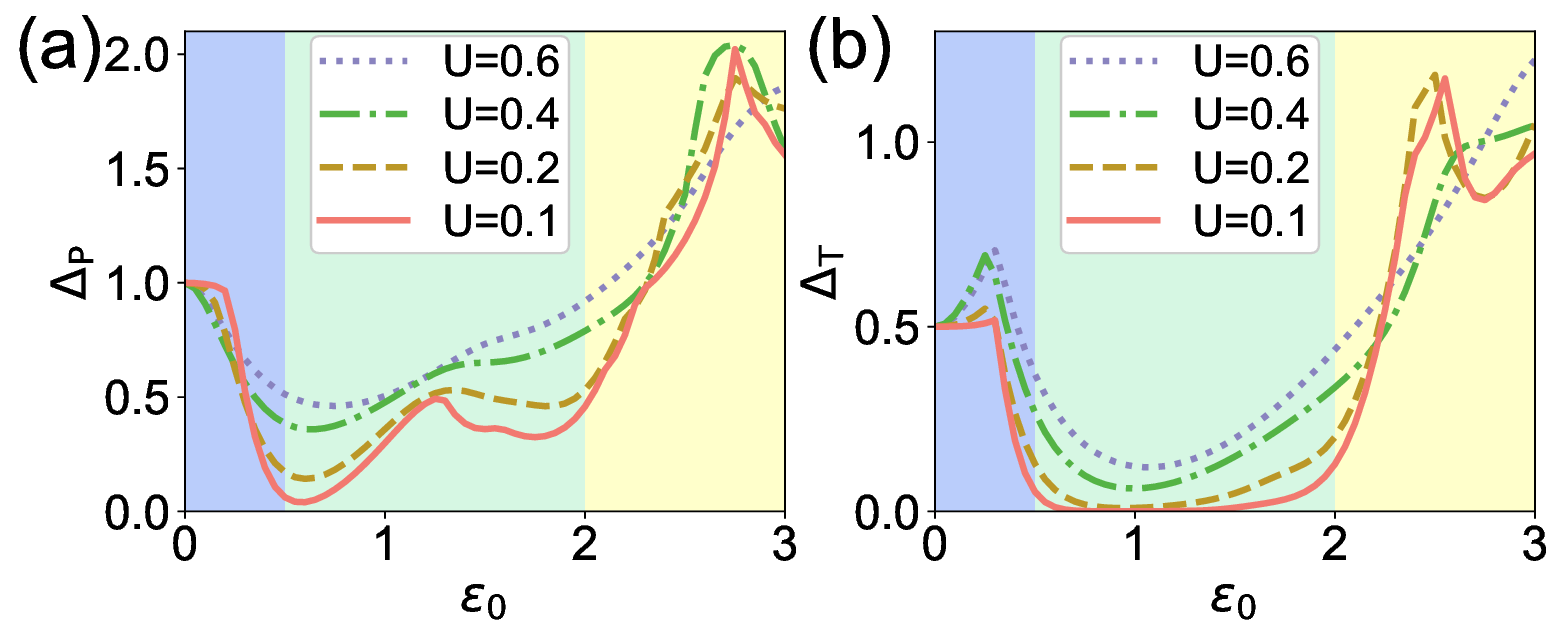}
	\caption{{Dissipative phase transition and dissipative discrete time crystal.} Two kinds of dissipative gaps $\Delta_{\mathrm{P}}$ ({a}) and $\Delta_{\mathrm{T}}$ ({b}) are shown in the figure (defined in the main text). The closing of $\Delta_{\mathrm{P}}$ and $\Delta_{\mathrm{T}}$ denotes the emergence of dissipative phase transitions and dissipative (discrete) time crystals.}
	\label{fig:d}
\end{figure}

{\it Dissipative phase transitions.}---In time-independent systems, the dissipative phase transition is characterized by the closing of the dissipative gap from both the real and imaginary axes. We find that such a definition also applies to the Floquet systems when we consider the dissipative gap in the effective Liouvillian spectrum. In Fig. \ref{fig:d}d, we plot the dissipative gap $\Delta_{\mathrm{P}}$ defined as the opposite of the maximum nonzero real eigenvalue, which is also the smallest absolute value of all nonzero eigenvalues in the considered parameter region. We can observe a clear closing of the dissipative gap at the first phase boundary near $\varepsilon=0.5$ in the thermodynamic limit. At the second phase boundary with the emergence of chaos, the dissipative gap also exhibits a decreasing tendency but remain away from zero. It denotes that the second phase transition may not be an exact dissipative phase transition but instead a crossover between two phases. The system only exhibits a weak symmetry $\mathbb{Z}_{2}$, which can only support a single phase transition through the spontaneous symmetry breaking. To observe a clear closing of the dissipative gap, we need to introduce additional symmetry, such as a strong symmetry. However, the crossover is enough to separate the regular and chaotic phases, as verified by the transitions of the qausi-steady observables.

Moreover, we investigate another dissipative gap $\Delta_{\mathrm{T}}$ defined as the opposite of the maximum real part of nonzero eigenvalues. As shown in Fig. \ref{fig:d}b, the gap $\Delta_{\mathrm{T}}$ closes in the second phase, the region with a regular attractor of double fixed points. The closing of $\Delta_{\mathrm{T}}$ denotes the emergence of pure imaginary eigenvalues and the dissipative (discrete) time crystal \cite{gong_discrete_2018,iemini_boundary_2018,li_time_2024,cabot_nonequilibrium_2024}. Here, the imaginary part of the considered nonzero eigenvalues with the maximum real part is $\pm\pi$. It means that the period of quantum evolution is doubled compared to the period of the Floquet drive. It is also consistent with the classical attractor, as the double fixed point in the Poincar{\' e} section also denotes a period doubling.

{\it Conclusion and discussion.}---
In conclusion, we establish a connection between the classical chaotic attractor and the Wigner distribution of a steady (quasi-steady) chaotic state in the semiclassical limit, as an analogy of the Berry conjecture in open systems. We corroborate our theory through a Floquet Kerr oscillator with Floquet squeezing drive. In this model, there is a discrete time crystal phase before the emergence of chaos. We show that the chaotic delocalization in the quasi-steady state is unique for the chaotic phase and can be quantitatively verified by the diverging entropy in the thermodynamic limit.

The most important prediction of our theory is that the chaotic steady state exhibits diverging entropy from the chaotic delocalization. For the system of bosonic modes, it is similar for the chaotic delocalization and the thermal delocalization, so the entropy scaling of the chaotic state can be well described by a thermal state. As a generalization, for spin systems with discrete degrees of freedom, the entropy of the chaotic steady state should satisfy a scaling of the total Hilbert space. This is verified recently in the many-body Bose-Hubbard trimer model \cite{rufo_quantum_2025}.

As shown in the supplementary materials \cite{sup}, the random matrix theory also works badly in the Floquet nonlinear oscillator. Our theory thus provides an important method to characterize the dissipative chaos in continuous variable systems. A direct application is to search for dissipative chaos in those quantum systems without classical analogy and beyond the region of non-Hermitian random matrix theory \cite{sup}. Moreover, the Floquet nonlinear model we considered here may be realized by four-wave mixing in an optical or superconducting cavity, and the chaotic delocalization in the phase space, as well as the resulting entropy diverging, provides characteristic phenomena in the experiment \cite{sup,micro_resonator,FWM2}.

\begin{acknowledgments}
	We thank Prof. Masahito Ueda at the University of Tokyo for helpful discussions. This work is supported by the National Key R\&D Program of China (Grant No. 2023YFA1407600), and the National Natural Science Foundation of China (NSFC) (Grants No. 123B2066, No. 12275145, No. 92050110, No. 91736106, No. 11674390, and No. 91836302).
\end{acknowledgments}

%

\end{document}